\documentclass[
    ,final            
  ]
  {aipproc}

\layoutstyle{6x9}

\usepackage{epsfig,amsmath,amssymb,verbatim,mathrsfs,subfigure}

\def\u1x{${U(1)_X}$}
\def\beq{\begin{equation}}
\def\eeq{\end{equation}}
\def\bea{\begin{eqnarray}}
\def\eea{\end{eqnarray}}
\def\bmat{\begin{pmatrix}}
\def\emat{\end{pmatrix}}

\def\mev{\,{\rm MeV}}
\def\gev{\,{\rm GeV}}

\def\zpri{{Z^{\prime}}}
\def\mzpri{{M_{Z^\prime}}}

\def\beq{\begin{equation}}
\def\eeq{\end{equation}}
\def\bea{\begin{eqnarray}}
\def\eea{\end{eqnarray}}
\def\bmat{\begin{pmatrix}}
\def\emat{\end{pmatrix}}

\def\mev{\,{\rm MeV}}
\def\gev{\,{\rm GeV}}

\def\zpri{{Z^{\prime}}}
\def\mzpri{{M_{Z^\prime}}}
\def\mh{{m_h}}
\def\mH{{m_H}}

\def\mpsi{M_\psi}

\def\brinv{$BR_{\rm inv}\ $}

\newcommand{ \slashchar }[1]{\setbox0=\hbox{$#1$}   
   \dimen0=\wd0                                     
   \setbox1=\hbox{/} \dimen1=\wd1                   
   \ifdim\dimen0>\dimen1                            
      \rlap{\hbox to \dimen0{\hfil/\hfil}}          
      #1                                            
   \else                                            
      \rlap{\hbox to \dimen1{\hfil$#1$\hfil}}       
      /                                             
   \fi}                                             %

\def\stacksymbols #1#2#3#4{\def\theguybelow{#2}
    \def\vp{\lower#3pt}
    \def\sp{\baselineskip0pt\lineskip#4pt}
    \mathrel{\mathpalette\intermediary#1}}

\def\intermediary#1#2{\vp\vbox{\sp
     \everycr={}\tabskip0pt
     \halign{$\mathsurround0pt#1\hfil##\hfil$\crcr#2\crcr
              \theguybelow\crcr}}}

\def\comment#1{}
\def\to{\rightarrow}

\newcommand{\nc}{\newcommand}
\nc{\LL}{L} \nc{\vv}{\tilde{v}} \nc{\ccdot}{\!\cdot\!}
\nc{\gsm}{G_{SM}}
\nc{\vfive}{\mathbf{5}\oplus\mathbf{\overline{5}}}
\nc{\vten}{\mathbf{10}\oplus\mathbf{\overline{10}}}
\nc{\zhol}{Z^{\rm hol}}

\begin{document}

\title{Hidden Sector Dark Matter and LHC Signatures}


\author{Shrihari Gopalakrishna}{
  address={Physics Department, Brookhaven National Laboratory, Upton, NY 11793. USA.}
  ,altaddress={New address: \\ The Institute of Mathematical Sciences, C.I.T Campus, Taramani, Chennai 600113. India.}
}

\begin{abstract}

We discuss the implications of a gauged Abelian hidden-sector  
communicating with the Standard Model (SM) fields via 
kinetic mixing with the SM hypercharge gauge field, or via the Higgs quartic interaction. 
We discuss signatures of the hidden-sector gauge boson at the LHC in the
four-lepton channel.
We show that a hidden-sector fermion can be a natural dark-matter 
candidate with the correct relic-density, 
discuss direct-detection prospects, 
and show how Higgs signatures may be altered at the LHC.

\end{abstract}

\maketitle



This paper summarises the analysis presented in Refs.~\cite{Gopalakrishna:2008dv} and 
\cite{Gopalakrishna:2009yz} and the reader is referred to these works for more details
and a fuller list of references.

\medskip\noindent{\it The theory:} 
The SM has two gauge invariant, flavor-neutral operators that are relevant (dimension $<4$): 
the hypercharge field-strength tensor $B_{\mu\nu}$ and
the SM Higgs mass operator $|\Phi_{SM}|^2$.  Hidden sector (i.e., non-SM states with no SM charge) 
abelian gauge bosons $X$ and Higgs bosons $\Phi_H$ can couple to these operators in a gauge invariant, 
renormalizable manner:
$X_{\mu\nu}B^{\mu\nu}$, and $|\Phi_H|^2|\Phi_{SM}|^2$.
In this letter we investigate the phenomenological implications of the
existence of these two operators.  

We consider an extra $U(1)_X$ factor in addition to the SM gauge group. 
Details are presented in Refs.~\cite{Gopalakrishna:2008dv} and \cite{Gopalakrishna:2009yz}, 
and related aspects can also be found in Refs.~\cite{u1xRel}.
We start by exploring the coupling of $X_\mu$ via kinetic mixing with $B_\mu$.
The kinetic energy terms of the \u1x gauge group are
${\cal L}^{KE}_X = -\frac{1}{4} \hat{X}_{\mu\nu} \hat{X}^{\mu\nu} + \frac{\chi}{2} \hat{X}_{\mu\nu} \hat{B}^{\mu\nu}$,
where we take the parameter $\chi \ll 1$ to be consistent with precision electroweak
constraints.  Hats on fields imply that gauge fields do not have canonically normalized kinetic terms.

We introduce a new Higgs boson $\Phi_{H}$ in addition to the usual SM Higgs boson
$\Phi_{SM}$.
Under $SU(2)_L \otimes U(1)_Y \otimes U(1)_X$ we take the representations
$\Phi_{SM}: (2, 1/2, 0)$ and $\Phi_{H}: (1, 0, q_X)$, with $q_X$ arbitrary.
$U(1)_X$ is broken spontaneously by $\left< \Phi_H \right> = \xi/\sqrt{2}$,
and electroweak symmetry is broken spontaneously as usual by
$\left< \Phi_{SM}\right> = (0,v/\sqrt{2})$.
The two real physical Higgs bosons $\phi_{SM}$ and $\phi_H$ mix after symmetry breaking,
and the mass eigenstates $h, H$ are related to the interaction states $\phi_{SM}, \phi_H$
by the sine of the mixing angle denoted as $s_h$ and the cosine as $c_h$.

\medskip\noindent{\it $X_\mu$ signals via $pp\to h\to XX\to \bar l l \bar l' l'$ :}
If the exotic gauge boson is sufficiently light, the lightest Higgs boson decays into a pair of 
them.  
The decay of the Higgs boson into two $X$ bosons is through Higgs boson mixing.  
The $X$ boson will then decay into SM fermions if there is even a tiny amount of kinetic mixing, 
which we assume to be the case.
We are particularly interested in leptonic final states, and we provide details of how 
$pp\to h\to XX\to \bar l l \bar l' l'$ is possible within this theoretical framework, 
and to explore the detectability of this channel at the Fermilab Tevatron and CERN LHC.

In presenting results in this section, we will choose
$\eta = 10^{-4}$, $\xi = 1$~TeV, and unless mentioned otherwise,
take $c_h^2 = 0.5$.
For illustration, we choose six benchmark points:
Points A~--~F with  
($M_h$, $\mzpri$) values in GeV given by 
(120, 5) ; (120, 50) ; (150, 5) ; (150, 50) ; (250, 5) ; (250, 50) respectively.
For these points we compute the differential distributions,
make cuts and find the significance at the Tevatron and LHC.
We make use of the narrow width approximation and analyse in succession: $pp\to h$
followed by $h\to \zpri\zpri$ followed by $\zpri\to \ell^+\ell^-$.


A $120\gev$ ($250\gev$) Higgs boson has total width of $\sim 10\mev$ ($\sim 2.1\gev$) when
$M_{Z'}=5\gev$ and $c_h^2=0.5$.
The $\zpri$ coupling to the SM sector is proportional to the tiny $\eta$,
making the width rather small, but these are the only modes kinematically
allowed for the $\zpri$ to decay into.
The $\zpri$ total width for $\eta=10^{-4}$ is $5.8\times 10^{-10}$, $2.7\times 10^{-9}$, $8.2\times 10^{-9}$ and $2.0\times 10^{-7}\gev$ for $M_{Z'}=5$, $20$, $50$ and $100\gev$ respectively.

The gluon fusion process $gg\to h$ is the largest production channel at the
Tevatron ($\sqrt{s}=1.96$~TeV) and the LHC ($\sqrt{s}=14$~TeV).
For instance, at the Tevatron, NLO $\sigma(gg \rightarrow h) = 0.85$~pb
for $M_h=120$ GeV while the sum of the other channels gives $0.33$~pb; the corresponding
cross-sections at the LHC are $40.25$~pb and $7.7$~pb
respectively. 
We include only gluon fusion computed at NLO using  HIGLU~\cite{NLOggh}.
We use MadGraph to obtain all matrix
elements, and generate event samples using MadEvent~\cite{MadGE}
with CTEQ6L1 PDF~\cite{Pumplin:2005rh}.


After applying suitable cuts (see Ref.~\cite{Gopalakrishna:2008dv}) to maximise signal while reducing background, 
we find the following cross-sections for points A -- F (in fb): 245, 44, 173, 57, 5.6, 2.2 respectively,
with the SM background (VV + hZZ) being 0.02, 0.02, 0.03, 0.03, 1.1, 1.1 respectively.  
We thus see that the prospect of discovering the $X_\mu$ in this channel is excellent at the LHC. 

\medskip\noindent{\it Hidden sector fermions:} 
We add to this theory two vector-like pairs of fermions ($\psi,\ \psi^c$) and ($\chi,\ \chi^c$)
that carry $U(1)_X$ charges but not any SM gauge quantum numbers.
Since there are no fermions charged under both the SM gauge group and $U(1)_X$, there
are no mixed anomalies. The vector-like nature makes the $U(1)_X$ anomaly cancellation
trivial.
We add the Lagrangian terms (written with Weyl spinors)
\beq
{\cal L} \supset 
-\lambda_s \Phi_H \psi \chi - \lambda_s^\prime \Phi_H^* \psi^c \chi^c 
-M_\psi \psi^c \psi - M_\chi \chi^c \chi + {\rm h.c.} 
\eeq
where the fermion covariant derivative terms are not shown,
and $q_\psi$ represents the $U(1)_X$ charge of $\psi$. 
We assume that the vector-like masses $M_\psi$ and $M_\chi$ are around the electroweak scale.

There is an accidental $Z_2$ symmetry under which $\psi,\ \psi^c,\ \chi,\ \chi^c$ are odd,
while $\Phi_H$ and all SM fields are even. This ensures the stability of the lightest $Z_2$ odd fermion,
which we will identify as the dark-matter candidate.

In addition to the vector-like masses, $U(1)_X$ breaking by $\left< \Phi_H \right> = \xi/\sqrt{2}$
implies the Dirac masses
$m_D \equiv \lambda_s \xi/\sqrt{2}$ and $m_D^\prime \equiv \lambda_s^\prime \xi/\sqrt{2}$.

We will explore the cosmological, direct-detection and collider implications
of the theory we have outlined. 
We will restrict ourselves to the lightest (and therefore stable) hidden sector fermion
(denoted as $\psi$ henceforth).
The relevant parameters are: $\mpsi$, $\kappa_{11}$ (the coupling of the hidden sector fermions to the hidden Higgs), 
$\kappa_{3\phi}$ (the Higgs cubic coupling), $s_h$ and $\mh$. 

\medskip\noindent{\it Relic density:}
$\psi\psi$ annihilations into the $W^+W^-$, $Z Z$, $h h$, $t\bar t$ final states will be important
if they are kinematically accessible, 
and if not, the dominant channel is into $b\bar b$. 
We compute the annihilation cross-section  
in the mass basis including $s$, $t$ and $u$-channel graphs.

\begin{figure}[ttt]
\includegraphics[angle=0,width=0.3\textwidth]{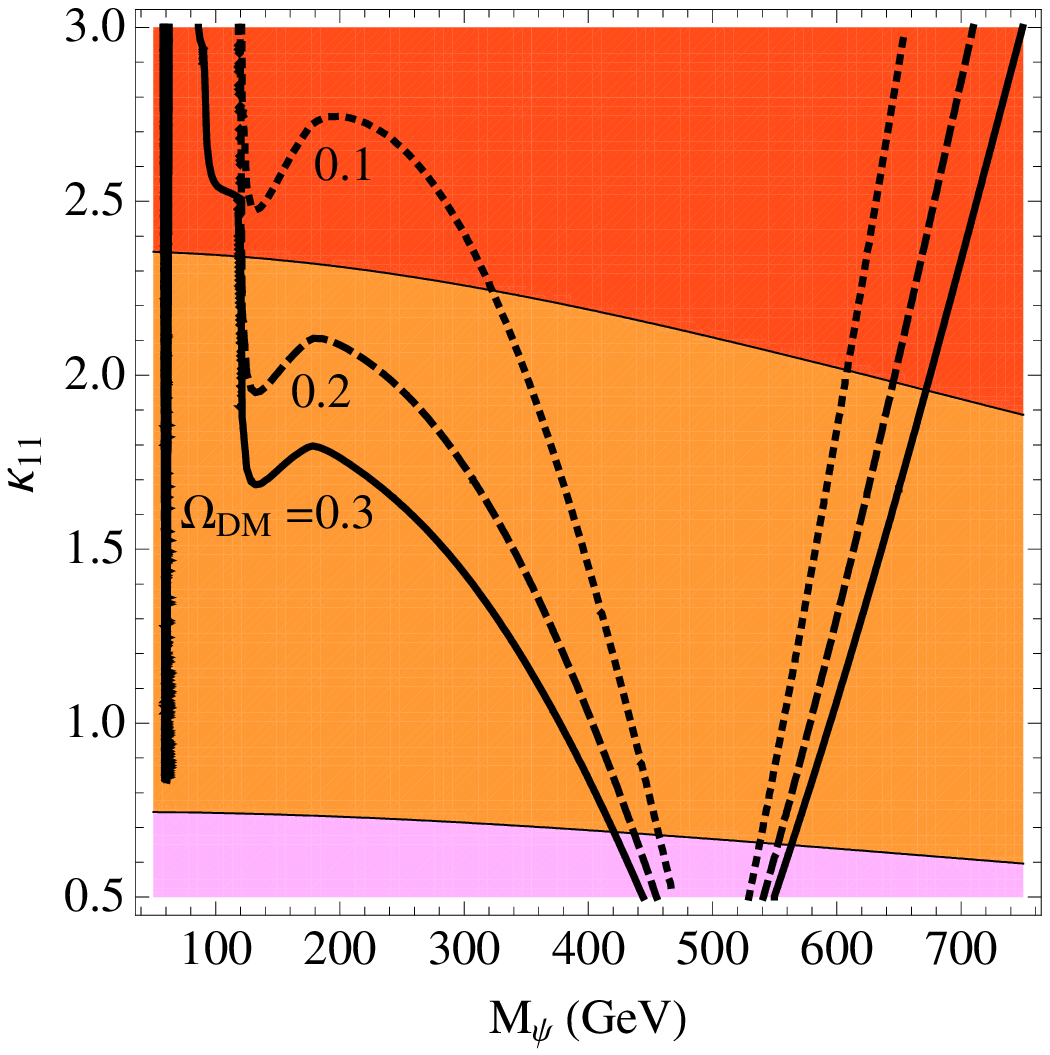}
\hspace*{0.25cm}
\includegraphics[angle=0,width=0.3\textwidth,height=0.2\textheight]{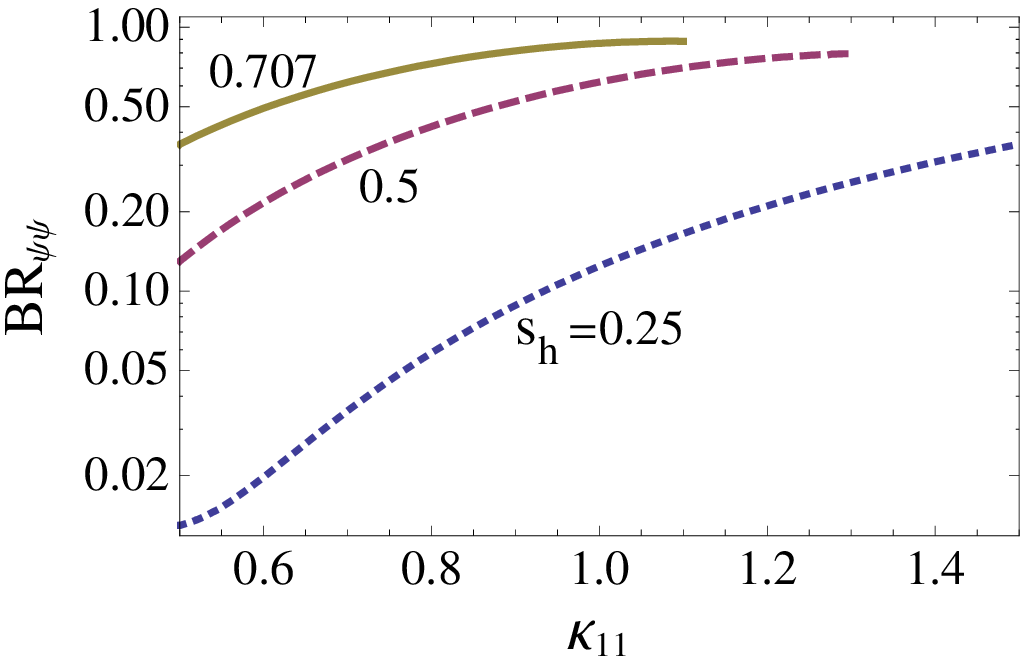}
\caption{
Left panel: Contours of ${\Omega_{dm}}_0 = 0.1,~0.2,~0.3$ (dot, dash, solid) in the 
$\mpsi$--$\kappa_{11}$ plane.
The direct-detection $\psi-N$ cross-section are shown as shaded regions:
$\sigma \gtrsim 10^{-43}~{\rm cm}^2$ (dark-shade)
is already excluded by experiments. 
$\sigma \gtrsim 10^{-44}~{\rm cm}^2$ (medium-shade),
and $\sigma \gtrsim 10^{-45}~{\rm cm}^2$ (light-shade), the latter two
will be probed in upcoming experiments.  
%
Right panel: The \brinv as a function of $\kappa_{11}$ for $\mh=120$~GeV
for $s_h=0.25,0.5,0.707$ (dotted, dashed, solid) with $\mpsi$ 
adjusted to give the correct dark matter relic density ($\Omega_0$).
\label{om0hBRinv.FIG}
}
\end{figure}

We show in Fig.~\ref{om0hBRinv.FIG} (left) the (0.1,~0.2,~0.3) contours of ${\Omega_{dm}}_0$ 
in the $\mpsi$--$\kappa_{11}$ plane, 
with the parameters not varied in the plots fixed at 
$\mpsi = 200~$GeV, $\mh = 120~$GeV, $s_h=0.25$, $\kappa_{11}=2.0$,  
$\kappa_{3\phi} = 1$, $\mH=1~$TeV, $\kappa_{H2h}=1$ and $\xi=1~$TeV.
This bench-mark point results in $\Omega_{dm} \approx 0.2$. 
We see that there exists regions of parameter space that are consistent with 
the present experimental observations.
In the region $\mh > 2 \mpsi$, the $h\to \psi\psi$ decay is allowed,
implying an invisibly decaying Higgs at a collider. This connection
will be explored in the following.
%

\medskip\noindent{\it Direct detection:}
Many experiments are underway currently to directly detect dark matter, and still
more are proposed to improve the sensitivity. 
In order to ascertain the prospects of directly observing $\psi$ in the \u1x framework we
are considering, we compute the elastic $\psi$-nucleon cross-section
due to the $t$-channel exchange of the Higgs boson. 
To illustrate, for $\kappa_{11}=2.0$, $s_h=0.25$, $\mpsi=200$~GeV, $\mh=120$~GeV, 
we find $\sigma \approx 1.9\times 10^{-16}~{\rm GeV}^{-2} = 7\times 10^{-44}~{\rm cm}^2$.
This is very interesting as the presently ongoing experiments~\cite{DirDetExptCite} 
are probing this range of cross-sections. 
With all other parameters fixed as above, as $\mh$ is increased to $350$~GeV, 
the direct-detection cross-section falls smoothly to about $10^{-45}~{\rm cm}^2$. 
In Fig.~\ref{om0hBRinv.FIG} (left)
we show the direct detection cross-section as shaded regions;
from the compilation in Ref.~\cite{DirDetExptCite}, 
the dark-shaded region ($\sigma \gtrsim 10^{-43}~{\rm cm}^2$) is excluded by 
present bounds from direct detection searches,
while the medium-shaded ($\sigma \gtrsim 10^{-44}~{\rm cm}^2$) and the 
light-shaded ($\sigma \gtrsim 10^{-45}~{\rm cm}^2$)
regions will be probed by upcoming experiments. 
We have defined our model into the package MicrOMEGAs~\cite{Belanger:2006is} 
and checked that our analytical results 
agree with the full numerical treatment reasonably well.

\medskip\noindent{\it Higgs Boson Decays:}
In addition to the usual SM decay modes,
if $\mpsi < \mh/2$, the decay $h \to \psi \bar\psi$ is kinematically allowed,
leading to an invisible decay mode for the Higgs boson.

We impose the requirement that the relic density should 
be in the experimentally measured range by scanning over 
$\mpsi \sim 60$~GeV, and show in 
Fig.~\ref{om0hBRinv.FIG} (right) the corresponding \brinv as a function of $\kappa_{11}$,
with $\kappa_{3\phi}=1.0$ and $\mH=1$~TeV held fixed.
%
%
We see that a significant \brinv is possible while giving the
required $\Omega_0$ and being consistent with present direct-detection limits,
with the general trend
of increasing \brinv for increasing $\kappa_{11}$ or $s_h$. 
Here we have shown only the points that satisfy the direct-detection 
cross-section $\sigma < 10^{-43}~{\rm cm}^2$,
to be consistent with current experimental results~\cite{DirDetExptCite}.
For a larger Higgs mass we find qualitatively similar invisible BR with
larger values of $\kappa_{11}$ preferred. 

\medskip\noindent{\it LHC Higgs phenomenology:} 
The discovery significance of the
light Higgs in the $gg\rightarrow h \rightarrow \gamma\gamma$, 
$gg\rightarrow h \rightarrow ZZ\rightarrow 4\ell$ and
$gg\rightarrow h \rightarrow WW\rightarrow 2\ell2\nu$
channels compared to those of a SM Higgs boson with the same mass is reduced 
appreciably, but we show that the prospects of discovering the Higgs 
via its invisible decay mode in the vector-boson-fusion channel becomes excellent. 

The vector-boson-fusion channel has been analysed in Ref.~\cite{Eboli:2000ze}, which we use to
obtain significances in the \u1x model by multiplying the signal cross-section
given there by \brinv\-$c_h^2$.
The backgrounds included there are QCD and EW $Z j j$ and $W j j$. 
We find in the \u1x model after suitable cuts (see Ref.~\cite{Gopalakrishna:2009yz}),
for $\mh = 120, 200, 300$~GeV, that we need for $5\,\sigma$ significance at the
LHC an integrated luminosity of $(0.44, 0.7, 1.3)\, / \, (BR_{inv}^{2}\, c_{h}^{4})$ $fb^{-1}$ respectively.
For example, for $\mh=120$~GeV, \brinv=0.75 and $s_h=0.5$, we would require a luminosity
of $1.4~{\rm fb}^{-1}$ for $5\,\sigma$ statistical significance. 
Alternatively, with $10~{\rm fb}^{-1}$, 
we can probe \brinv down to about $26$\,\% at $5\,\sigma$.
We thus see that in this channel, the
prospect of discovering an invisibly decaying Higgs boson in the \u1x scenario is excellent.


{\it Acknowledgments: }
We thank the organisers of SUSY09 for a very nice conference. 
SG was supported in part by the DOE grant DE-AC02-98CH10886 (BNL).

\end{document}